\documentclass[lettersize,journal]{IEEEtran}
\usepackage{amsmath,amsfonts}
\usepackage{algorithmic}
\usepackage{algorithm}
\usepackage{array}
\usepackage[caption=false,font=normalsize,labelfont=sf,textfont=sf]{subfig}
\usepackage{textcomp}
\usepackage{stfloats}
\usepackage{url}
\usepackage{verbatim}
\usepackage{graphicx}
\usepackage{cite}
\usepackage{xcolor}
\begin{document}

\title{TiRE-GAN: Task-Incentivized Generative Learning for Radiomap Estimation}

\author{Yueling Zhou, Achintha Wijesinghe, Yibo Ma, Songyang Zhang, ~\IEEEmembership{Member, IEEE}, and Zhi Ding, ~\IEEEmembership{Fellow, IEEE}

\thanks{This work was supported in part by the National Science Foundation under Grants No. 2425811, No. 2431452 and No. 2332760. (Corresponding author: Songyang Zhang).}
\thanks{Y. Zhou and S. Zhang are with Department of Electrical and Computer Engineering, University of Louisiana at Lafayette, Lafayette, LA, 70503. (E-mail: \{yueling.zhou1, songyang.zhang\}@louisiana.edu).}
\thanks{A. Wijesinghe, Y. Ma and Z. Ding are with Department of Electrical and Computer Engineering, University of California, Davis, CA, 95616. (E-mail: \{achwijesinghe, boma, zding\}@ucdavis.edu).}
}

\markboth{Journal of \LaTeX\ Class Files,~Vol.~14, No.~8, August~2021}%
{Shell \MakeLowercase{\textit{et al.}}: A Sample Article Using IEEEtran.cls for IEEE Journals}

\IEEEpubid{0000--0000/00\$00.00~\copyright~2021 IEEE}

\maketitle

\begin{abstract}
To characterize radio frequency (RF) signal power distribution in wireless communication systems, the radiomap is a useful tool for resource allocation and network management. Usually, a dense radiomap is reconstructed from sparse observations collected by deployed sensors or mobile devices. To leverage both physical principles of radio propagation models and data statistics from sparse observations, this work introduces a novel task-incentivized generative learning model, namely TiRE-GAN, for radiomap estimation. Specifically, we first introduce a radio depth map to capture the overall pattern of radio propagation and shadowing effects, following which a task-driven incentive network is proposed to provide feedback for radiomap compensation depending on downstream tasks. Our experimental results demonstrate the power of the radio depth map to capture radio propagation information, and the efficiency of the proposed TiRE-GAN for radiomap estimation.
\end{abstract}

\begin{IEEEkeywords}
Radiomap estimation, generative adversarial networks, radio propagation model
\end{IEEEkeywords}

\section{Introduction}
Rapid growth of sensor networks and edge computing continue to advance the development of next-generation wireless communications, stimulating many novel technologies, such as vehicle-to-everything \cite{c1} and integrated sensing and communications \cite{c2}. All
these applications require an accurate description of radio
frequency (RF) spectrum coverage and efficient assessment of
the radio environment, leading to the concept of radiomaps \cite{SB2019}. Radiomaps describe RF power spectral density (PSD) across various locations, frequencies, and time, offering rich information of spectrum coverage for the spectrum management applications. Usually, dense radiomaps are reconstructed from sparse observations collected by mobile sensors. 
To fully leverage the power of radiomaps
in spectrum management, 
efficient radiomap estimation (RME) from sparse observations
has emerged as an urgent challenge.

Existing methods of radiomap estimation can be categorized into model-based or learning-based approaches, each bearing its distinct constraints. Model-based approaches, such as the log-distance path loss (LDPL) models \cite{ldpl}, inverse distance weighted interpolation (IDW) \cite{idw}, and thin-plate splines kernels \cite{kernel}, usually assume a specific radio propagation model, with which unknown
parameters are estimated by minimizing the error between
observed samples and predictions. However, in practical applications, these physics models are often unknown, making model selection challenging.
Moreover, the PSD distribution sometimes is more sensitive to surrounding
obstacles rather than following a strict empirical model, especially in complicated environments. Thus, the efficient extraction of
radio model information from realistic data samples remains
an open question.

With the development of artificial intelligence (AI), recent interests of RME focus on learning-based methods \cite{DR2022}. In \cite{RL2021}, a fast radiomap estimation is proposed based on Unet. Another work utilizing deep convolutional neural networks
is \cite{auto2022}, where an auto-encoder structure is presented for radiomap estimation. Other learning-based methods also include conditional generative adversarial nets (cGAN) \cite{SZ2023,c00} and transformers \cite{transformer2021}. However, these learning-based methods require high-quality, abundant data, which is usually inaccessible in practical applications. Although some hybrid methods \cite{rmeinpainting, mbest} attempt to combine radio models with machine learning approaches, the efficient model selection and task-specific designed RME, such as for outage detection and network planning, remain critical concerns.

\IEEEpubidadjcol
To address the aforementioned problems, we propose a \textbf{T}ask-\textbf{i}ncentivized \textbf{R}adiomap \textbf{E}stimation using \textbf{G}enerative \textbf{A}dversarial \textbf{N}etworks (TiRE-GAN), leveraging the synergies of radio propagation
model and generative AI. Specifically, a radio depth map is
introduced as an additional input channel of the generator
to capture the overall radio propagation information and
shadowing effects. Then, a task-incentivized block is proposed
to provide feedback for radiomap generation to compensate
for the estimated PSD based on downstream tasks. Our
contributions can be summarized as follows:

We introduce a novel radio depth map (RDM) as an input to the cGAN, capturing shadowing and radio propagation effects to enhance the learning process with physical insights. We develop a physics-conditioned cGAN with a task-incentivized network, which improves RME performance by compensating radiomaps based on downstream tasks. Experiments demonstrate the effectiveness of the proposed RDM in capturing radio propagation behavior and the efficiency of TiRE-GAN. 
\IEEEpubidadjcol


\section{Problem Description}
\begin{figure*}[t]
    \centering
\includegraphics[width=0.75\textwidth]{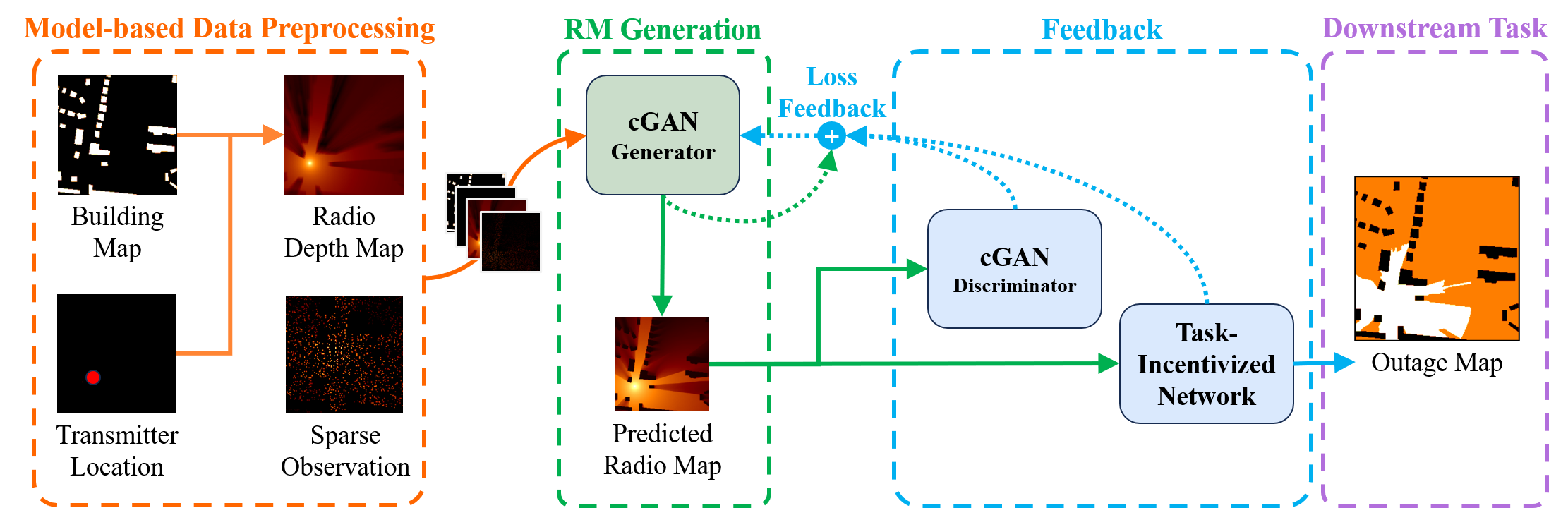}
    \caption{Overall Structure of TiRE-GAN consisting of 4 parts: 1) The model-based data preprocessing module utilizes the input observed information (i.e. building map and transmitter location) to generate the radio depth map and then stacks the four layers together as input; 2) The radiomap (RM) generation module is in a cGAN generator structure, which learn and produce the prediction of radiomap; 3) The feedback module consists of 2 parts: a cGAN discriminator and a task-incentivized network. Both modules produce a loss term as feedback to improve the generator; 4) The Downstream task that provides the additional loss term for training guidance.}
    \label{fig:phycGAN}
    \vspace{-5mm}
\end{figure*}

In this section, we first introduce our problem formulation. 
In this work,
the radiomap estimation is formulated as a sparse sensing problem, where a dense radiomap $\hat{M}_R$ is reconstructed given the landscape information of the surrounding environment, sparse observations collected from sensors, and locations of transmitters, formulated by the following equation:
\begin{equation}\label{problem description}
    \hat{M}_R = g(M_U, M_T, M_S, \mathcal{X}), 
\end{equation}
{where $g(\cdot)$ denotes the RME neural network. Here, $M_U$, $M_T$, and $M_S$ represent the landscape map, the encoded transmitter location, and the sampled radiomap, respectively. $\mathcal{X}=f_{physical}(M_U, M_T)$ is a function embedding the physical principles of radio propagation behavior. In this work, radio depth map is applied to capture the $f_{physcal}(\cdot)$, which will be introduced with details in Section \ref{RDM}.}

Different from the interpolation formulation in \cite{ldpl} and \cite{idw}, where the unobserved PSD values are estimated only using the observed samples in the same radiomap, we follow the classic prediction formulation and assume a set of training pairs of $(M_R, \mathcal{F})$ with $\mathcal{F}=\{M_U,M_T,M_S,\mathcal{X}\}$ to train the mapping function $g(\cdot)$ by
\vspace{-1mm}
\begin{equation}
    \min_{g(\cdot)} \quad L(M_R, \hat{M}_R),
    \vspace{-1mm}
\end{equation}
where $L(\cdot)$ refers to the designed loss function.

In this work, our objectives are to: 1) design an effective mapping function $g(\cdot)$ by integrating the cGAN with a novel task-incentivized network; and 2) design a suitable loss function $L(\cdot)$ to embed the physical principles of radio propagation. We introduce the details of our proposed TiRE-GAN in Section \ref{sec:method}.

\vspace{-3mm}
\section{Methods} \label{sec:method}
\subsection{Overall Structure of TiRE-GAN Framework}
We now introduce the overall structure of our proposed TiRE-GAN. As shown in Fig. \ref{fig:phycGAN}, the architecture of the proposed framework can be structured into three components: 1) model-based data preprocessing; 2) radiomap generation; and 3) task-incentivized feedback. 
The training process operates as follows. First, the urban landscape map and transmitter location data are processed to construct a physics-embedded Radio Depth Map (RDM), detailed further in Section \ref{RDM}, after which it is fed into the cGAN generator together with
the urban map, transmitter locations and sparse observations. During training, a pre-trained task-incentivized network is integrated to provide feedback by embedding physics-based constraints and rewards from downstream tasks. Each module will be introduced as follows.

\vspace{-4mm}
\subsection{Model-Based Data Preprocessing}
The data preprocessing module consists of two parts: 1) process the observed features $\{M_U, M_T, M_S\}$ into aligned inputs of cGAN; and 2) generate a radio depth map embedding the radio propagation behavior.
\subsubsection{Observed Data Preprocessing}
Suppose that the radiomap is formatted in a $N\times N$ grid, i.e., $M_R\in\mathbb{R}^{N\times N}$. The landscape map of buildings is encoded as a binary segmented image $M_U\in\mathbb{R}^{N\times N}$, where building areas are marked by 1 and open areas are annotated by 0. Similar to the building map, we encode the transmitter locations into a binary image $M_T\in\mathbb{R}^{N\times N}$, where the transmitter locations are marked as 1; otherwise, the pixel values are zero.
The sparse observations can be viewed as a grayscale image $M_S\in\mathbb{R}^{N\times N}$ via zero-padding, where we use the observed PSD the as corresponding pixel values; otherwise, the pixel value is zero.
For convenience, we first use a threshold to eliminate the extremely low-value observations, and then normalize the PSD value into $0\sim1$ for the observed locations. Since linear normalization is applied, we can easily re-scale the estimated radiomap back to the original data domain. For the missing values in $M_S$, we set them as zeros. Since a convolutional neural network is applied for the cGAN structure, zero-padding does not impact the final RME.

\subsubsection{Radio Depth Map}\label{RDM}
{In Eq. \eqref{problem description}, we included an $\mathcal{X}$ of additional extracted features for RME.} To embed the radio propagation information, we introduce the Radio Depth Map (RDM), {$\mathcal{X}=M_D\in\mathbb{R}^{N\times N}$ as a novel model-based attribute, derived from the distribution of buildings and the spatial locations of transmitters.} Similar to a depth map in computer vision which provides information on the distances of surfaces from a specific viewpoint, the RDM is formulated as:
\begin{equation}\label{depthmap}
M_D(\xi, \eta)=n\left(\sum_{t=1}^{N_T} P_t(\xi, \eta) \cdot B_t(\xi, \eta)\right),
\end{equation}
where $(\xi, \eta)$ denotes the two-dimensional coordinates of the target location, $N_T$ is the number of transmitters, $P_t(\xi, \eta)$ denotes the distance-based path loss gain from the $t$-th transmitter, $n(\cdot)$ is the max normalization function to re-scale the maximal value in the RDM as $1$, and the $B_t(\xi, \eta)$ capture the impact of buildings from the viewpoint of $t$-th transmitter.

Let $C(\ell,\ell_t)$ denote the set of pixels in the direct line of sight between the target location $\ell=(\xi, \eta)$ and location $\ell_t=(\xi_t,\eta_t)$ of the $t$-th transmitter. The shadowing term impacted by buildings is calculated by
\vspace{-1mm}
\begin{equation}
B_t(\xi, \eta)=\frac{\sum_{i \in C(\ell,\ell_t)} (1 - M_U(i))}{\sum_{i \in C(\ell,\ell_t)} 1},
\vspace{-1mm}
\end{equation}
which characterizes the ratio of non-building pixels between $\ell$ and $\ell_t$. $B_t(\xi, \eta)$ is larger and the shadowing effect is smaller if the radio propagates through fewer buildings.

To capture the path loss $P_t(\xi, \eta)$, we employ the Inverse Distance Weighting (IDW) model, calculated by
\begin{equation}\label{IDW}
P_{t}(\xi, \eta) = d(\ell,\ell_t)^{-\lambda}, \ \lambda > 0,
\end{equation}
where $d(\ell,\ell_t)$ denotes the Euclidean distance between the target $\ell$ and the $t$-th transmitter. The parameter $\lambda$ governs the rate, at which signal strength diminishes with increasing distance, necessitating a positive value to ensure physical plausibility in the depth map. An exemplary radio depth map is shown in the orange block of Fig. \ref{fig:phycGAN}.

\vspace{-4mm}
\subsection{cGAN Network}\label{cGAN}

In this work, we employ a cGAN \cite{MM2014} as the foundational architecture, which consists of two principal components: a generator $G$ and a discriminator $D$. The generator employs a ResNet-based architecture, consisting of 6 ResNet blocks. 
The discriminator uses a PatchGAN structure, comprising 4 convolutional blocks
, and ends with a Sigmoid activation to classify the input as real or generated.
{In our radiomap estimation, the conditional features consist of four channels as processed in the model-based data preprocessing module, 
i.e.,  $\mathcal{F}=\{M_U, M_T, M_S, M_D\}$ as shown in Fig. \ref{fig:phycGAN}.}
{The additional information $\mathbf{I}$ can be designed as a function of the region features, i.e., $\mathbf{I}=h(\mathcal{F})$ or some determined labels, such as surrounding landscape information}
\cite{cgan}. Then, the basic cGAN can be formulated as
\begin{align}
    \min_G\max_D \quad V(D,G)= \mathbb{E}_{y,I\sim p_{\rm data}(\bf{y},\bf{I})}[\log D(\bf{y,I})] +\nonumber\\
    \mathbb{E}_{f\sim p_{\rm data}(\bf{\mathcal{F}})}[\log (1-D(G(\bf{\mathcal{F}}),\mathbf{I}))],
\end{align}
where $p_{\rm data}(\bf{y,I})$ is the joint distribution with the prior knowledge $\mathbf{I}$, and $\mathbf{y}$ is the true radiomap. {$p_{\rm data}(\mathbf{\mathcal{F}})$ denotes the distribution of input features used for cGAN training. Interested readers could refer to \cite{cgan,SZ2023} for a detailed introduction to cGAN structure.}
Generally, the optimization of a min-max problem in cGAN can be split into two sub-problems, and train $D$ and $G$ iteratively. Specifically, the loss function to optimize $D$ is
   $ L_D=-V(D,G)$,
and the general loss function to optimize $G$ is 
\begin{equation}\label{gan_obj}
    L_G=\mathbb{E}_{x\sim p_{\rm data}(\bf{\mathcal{F}})}[\log (1-D(G(\bf{\mathcal{F}}),\mathbf{I}))].
\end{equation}

To leverage both radio propagation model information and data statistics from observed samples, in this work, 
we will customize the loss function of the generator as
\begin{equation}\label{eq:loss}
    L=L_G+\alpha L_{MSE}+\beta L_R.
\end{equation}
Here, $L_{MSE}$ describes the error of RME by Mean Squared Error (MSE), i.e.,
 $   L_{MSE}(\mathbf{y}, \hat{\mathbf{y}}) = \frac{1}{K} \sum_{i=1}^{K} (y_i - \hat{y}_i)^2$,
where $K$ is the total number of pixels, {$\mathbf{y}$ and $\hat{\mathbf{y}}$ are the ground truth and the predicted radiomap, while $y_i$ and $\hat{y}_i$ are the true and predicted pixel value of the $i$th pixel}. $L_R$ is a task-incentivized regularization term from the feedback module, which will be introduced in Section \ref{sec:fed}.

\vspace{-4mm}
\subsection{Feedback Design}\label{sec:fed}
The feedback module consists of two components: 1) a traditional cGAN discriminator; and 2) a task-incentivized network (TIN) based on downstream tasks. 

\subsubsection{cGAN Discriminator}
Conversely to the cGAN generator, the discriminator aims to detect whether the generated data is fake or not. Here, we consider three-channels as the input of the discriminator, where two are expanded image-sized one-hot encodings representing the binary truthfulness of the input, and the third channel is the generated radiomap. The discriminator outputs a scalar value, indicating the likelihood of whether the provided radiomap is genuine. The discriminator is trained by the binary cross entropy (BCE) loss as
\vspace{-2mm}
\begin{equation}
    L_D(\mathbf{\delta}, \hat{\mathbf{\delta}}) = -\frac{1}{K} \sum_{i=1}^{K} \left[ \delta_i \log(\hat{\delta}_i) + (1 - \delta_i) \log(1 - \hat{\delta}_i) \right],
    \vspace{-4mm}
\end{equation}
where $\delta$ and $\hat{\delta}$ are the true and predicted values. $K$ is the number of data points in the batch.
\begin{figure}
    \centering
    \includegraphics[width=0.4\textwidth]{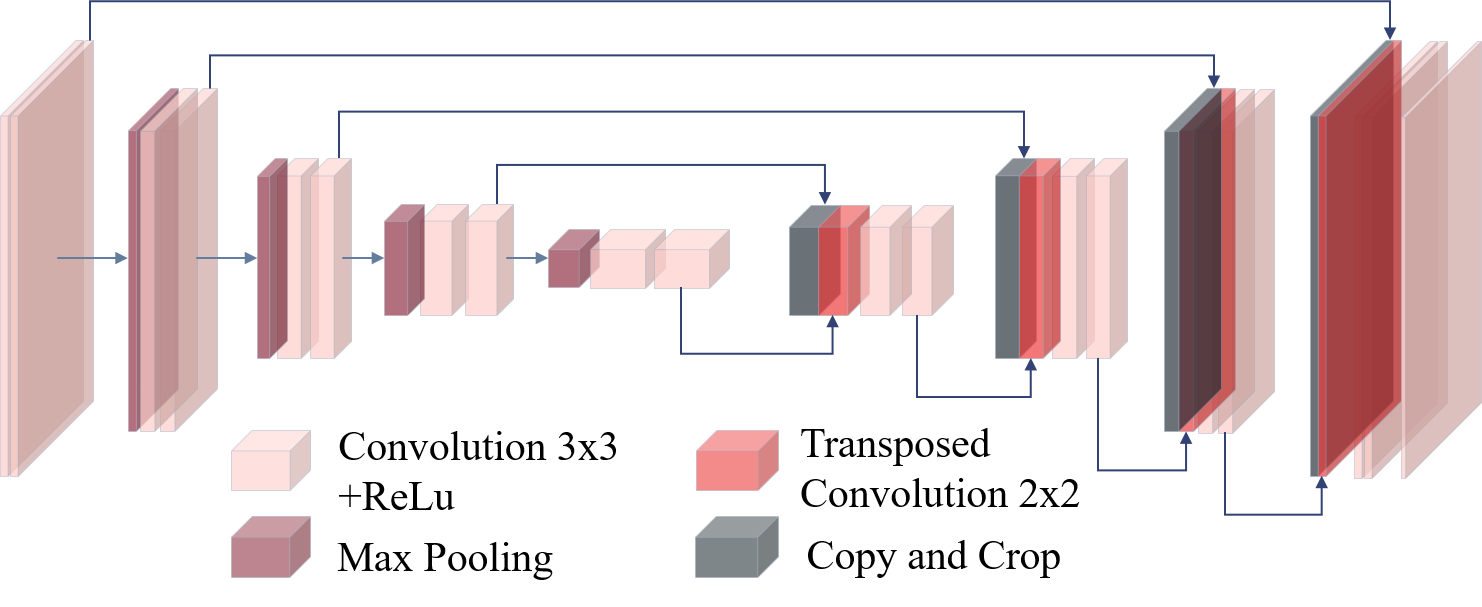}
    \caption{Task-incentivized network (TIN): based on Unet, TIN contains 4 encoding blocks and 4 decoding blocks. Each encoding block uses 3 convolution layers and a pooling layer. Each decoding block consists of one concatenation layer, a transpose convolution layer, and 3 convolution layers. }
    \label{fig:UNet}
    \vspace{-5mm}
\end{figure}
\subsubsection{Task-Incentivized Network}
The goal of TiRE-GAN extends beyond capturing radio signal strength at specific locations; it shall ultimately support spectrum planning and network management. For instance, radiomaps can highlight service outages by setting a failure threshold. To enhance radiomap estimations for downstream tasks, we introduce a TIN module, as shown in Fig. \ref{fig:phycGAN}.

We use outage detection as an exemplary TIN, employing a UNet architecture for semantic segmentation of radio power outages as shown in Fig. \ref{fig:UNet}. The input is a radiomap, and the output outage map indicates signal coverage. Trained on real radio coverage and outage maps, the TIN accurately predicts outage areas. The TIN operates within
the cGAN framework as a pre-trained component, for which the generator's output serves as input. The TIN then contributes an additional loss term, $L_R$, guiding cGAN to produce radiomaps that not only maintain precision but also address real-world concerns relevant to downstream tasks.

The TIN provides a task-incentivized regularization term $L_R$ in the loss function of the generator, which is calculated by the pre-trained TIN using the MSE loss to measure the error between the predicted and the ground truth outage maps, i.e.,
\vspace{-2mm}
\begin{equation}
    L_{R}(z, \hat{z}) = \frac{1}{N_p} \sum_{i=1}^{N_p} (z_i - \hat{z}_i)^2,
    \vspace{-2mm}
\end{equation}
where the $z$ and $\hat{z}$ are the ground truth and the predicted outage map, while $z_i$ and $\hat{z}_i$ are the true and predicted pixel values of the $i$th pixel. $N_p$ is the number of pixels.

Note that our TIN can be integrated with any RM-assisted applications. Some typical examples include ray tracing or wireless localization.

\begin{figure*}[t]
    \centering
    \includegraphics[width=0.8\textwidth]{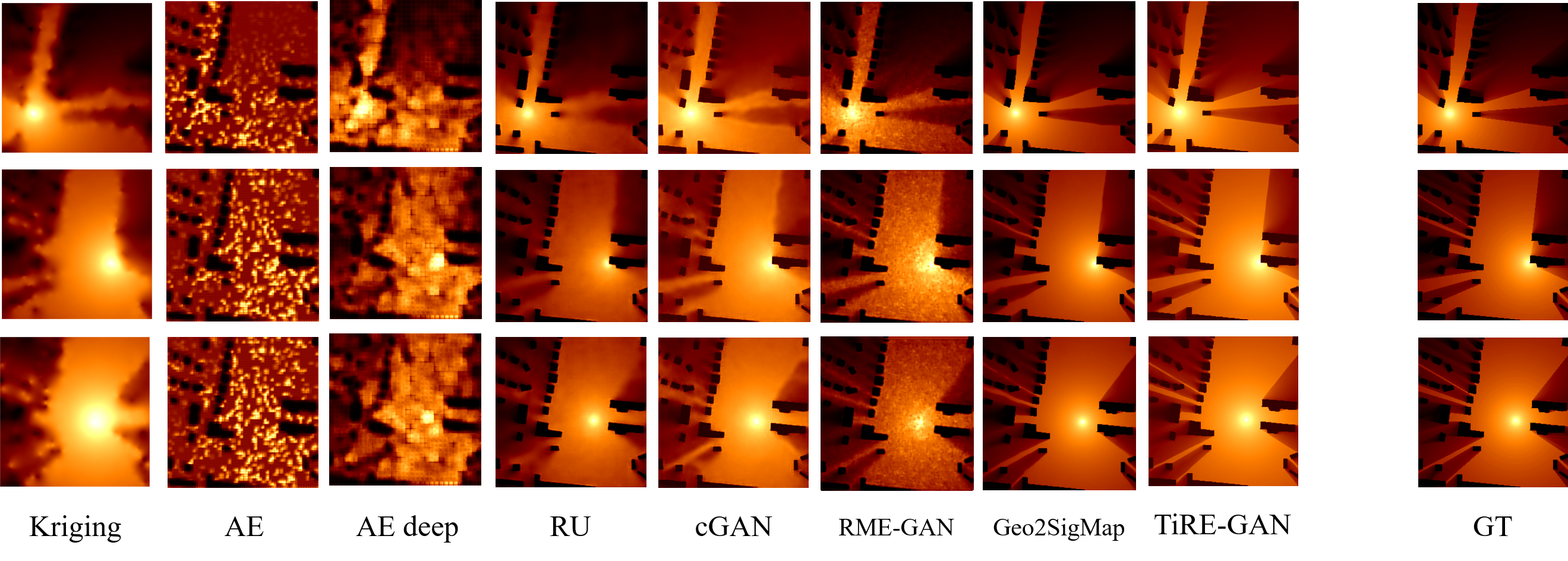}
    \vspace{-3mm}
    \caption{Performance comparison (sr=1) between Kriging, 
    AutoEncoder (AE), RadioUNet (RU), conditional GAN (cGAN), RME-GAN, Geo2SigMap, our TiRE-GAN, and the ground truth (GT) on 3 radiomap examples. TiRE-GAN achieves the best radio propagation and the shadowing effect prediction. }
    \label{fig:perf3}
    \vspace{-3mm}
\end{figure*}

\vspace{-3mm}
\section{Experiments}
\subsection{Experiment Setup}
\subsubsection{Data}
In this work, we utilize the RadiomapSeer dataset \cite{RL2021} for evaluation, which contain an extensive set of 700 maps from a variety of metropolitan areas, including Ankara, Berlin, Glasgow, Ljubljana, London, and Tel Aviv. Each map delineates a quadrant of $256\times256$ square meters, at a granularity of $256\times256$ pixels. The edifices are encoded by assigning a pixel intensity of one, whereas open spaces are represented by a zero value. Furthermore, the precise locations of transmitters are encoded with a pixel intensity of one superimposed on a null background, facilitating accurate localization.

\subsubsection{Comparison}
For performance assessment, we compare TiRE-GAN with both state-of-the-art (SOTA) model-based methods, including {Kriging} \cite{KS2017}, and data-driven approaches, including 
{AutoEncoder (AE)} \cite{auto2022}, {RadioUnet (RU)} \cite{RL2021}, conventional {cGAN} \cite{MM2014}, {RME-GAN} \cite{SZ2023}, and {Geo2SigMap} \cite{YL2024}. For interpolation-based method, we reconstruct the radiomap with observations in the same radiomap. For prediction-based method, we split the dataset as 4000/800/800 (70\%/15\%/15\%) for training/validation/testing sets. Since RME-GAN is designed for non-uniformly distributed samples, we remove the corresponding loss terms to ensure fair comparison. In our TiRE-GAN, we set $\alpha=1$ and $\beta=0.1$ in Eq. \eqref{eq:loss}.

\vspace{-4mm}
\subsection{Overall performance}

We first present the overall performance of different radiomap estimations. We evaluate the performance based on the normalized radiomaps via MSE denoted by
 $\text{MSE}(\mathbf{y}, \hat{\mathbf{y}}) = \frac{1}{N_p} \sum_{i=1}^{N_p} (y_i - \hat{y}_i)^2$,
and normalized MSE (NMSE) calculated by
 $    \text{NMSE}(\mathbf{y}, \hat{\mathbf{y}}) = \frac{1}{N_p} \sum_{i=1}^{N_p} \frac{(y_i - \hat{y}_i)^2}{y_i^2}$,
where $N_p = 256^2$. Linear normalization makes it easy to re-scale the estimated radiomap back to the original data domain.
From the results given in Table \ref{tab:NMSEandMSE}, our proposed TiRE-GAN always achieves superior performance against all other approaches. Different from RME-GAN and purely data-driven models, our radio depth map can capture both pathloss and shadowing efficiently, providing efficient guidance of cGAN training. With the assistance of radio depth map, our method can achieve a very high reconstruction accuracy even for very low sampling ratio and few training samples.
Fig. \ref{fig:perf3} presents the visualization results of different methods for three radiomap examples at a 1\% sample ratio. Advanced learning methods, such as RadioUnet, cGAN, and our TiRE-GAN, outperform Kringing 
and AE in capturing overall radio propagation patterns. 
Notably, due to the model-based interpolation block and complex loss design, RME-GAN cannot efficiently function at very low sampling ratio below 0.1\%. 
TiRE-GAN shows clearer boundaries for shadowing effects resulted from the TIN feedback. Both the visualization and numerical results validate the effectiveness of our TiRE-GAN in RME with insufficient training samples.

\begin{table*}[htbp] 
    \centering 
    \caption{Test NMSE ($\times 10^{-2}$)/MSE ($\times 10^{-4}$) Comparison} 
    \vspace{-2mm}
    \begin{tabular}{c|cccccccc} 
    \hline 
        sr (\%) & 0.03 & 0.1 & 0.5 & 1 & 3 & 5 & 10 & Average ($0.5\sim 10$)\\ 
    \hline 
        Kriging & 24.56/188.75 & 18.92/141.37 & 14.24/110.79 & 12.96/101.54 & 14.71/91.99 & 11.35/89.04 & 10.96/86.02 & 12.84/95.88 \\
        AE & 54.76/230.65 & 53.84/226.61 & 45.66/194.47 & 43.33/186.57 & 26.00/113.04 & 20.71/87.72 & 14.32/65.01 & 30.00/129.36 \\
        AE Deep & 50.65/209.26 & 48.81/201.40 & 21.42/91.95 & 16.25/70.00 & 6.82/26.81 & 6.55/26.43 & 5.81/23.47 & 11.22/47.73 \\
        RU & 6.54/24.64 & 5.11/19.38 & 2.63/10.02 & 2.22/8.46 & 1.81/7.08 & 1.51/5.96 & 1.27/4.99 & 1.89/7.30 \\
        cGAN & 11.57/45.88 & 5.48/22.09 & 2.13/8.46 & 1.83/7.25 & 1.30/5.07 & 1.18/4.64 & 1.02/4.05 & 1.49/5.89 \\
        RME-GAN & -/- & -/- & 0.77/7.42 & 0.68/5.70 & 0.64/6.15 & 0.53/5.08 & 0.47/4.51 & 0.60/5.77 \\
       Geo2SigMap & 2.19/8.67 & \textbf{1.23}/4.73 & 1.12/4.31 & 0.75/2.89 & 0.79/2.93 & 0.79/2.94 & 0.95/3.75 & 0.93/3.36 \\
        TiRE-GAN & \textbf{1.98}/\textbf{7.06} & 1.26/\textbf{4.65} & \textbf{0.72}/\textbf{2.77} & \textbf{0.61}/\textbf{2.35} & \textbf{0.52}/\textbf{2.04} & \textbf{0.51}/\textbf{2.00} & \textbf{0.46}/\textbf{1.81} & \textbf{0.55}/\textbf{2.19} \\
    \hline 
    \end{tabular} 
    \label{tab:NMSEandMSE} 
    \vspace{-2mm} 
\end{table*}

 \vspace{-3mm}
\subsection{Robustness Against Noise}
\begin{figure}
    \centering
    \includegraphics[width=0.49\textwidth]{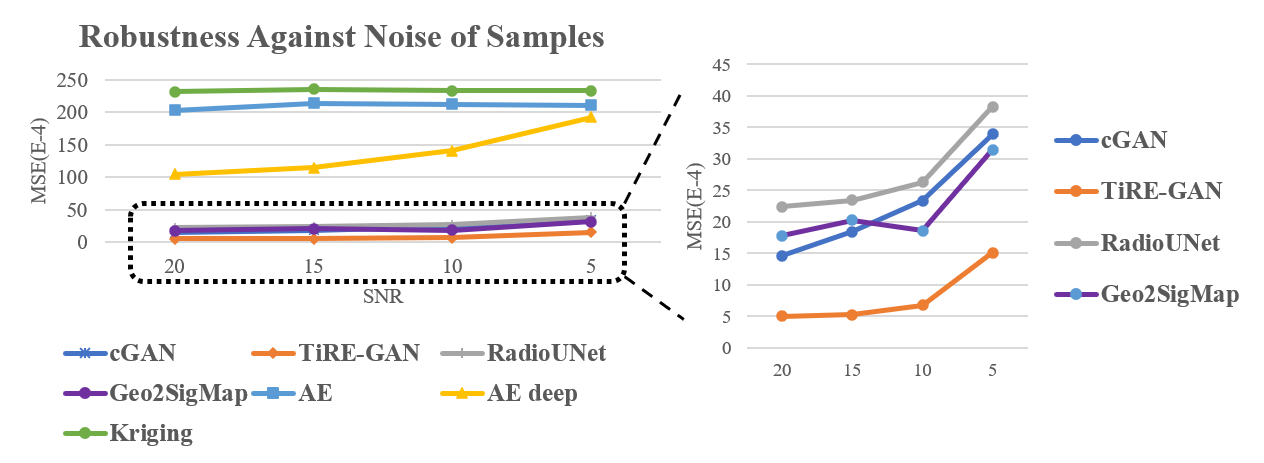}
    \caption{The MSE ($\times 10^{-4}$) comparison with different SNRs. 
    }
    \label{fig:rob}
\end{figure}

To assess the robustness, we include additive Gaussian noise to each sampled radiomap value, replacing the original sparse observation channel with a noisy input pattern. To highlight the effectiveness of the proposed radio depth map and TIN blocks, we compared different methods across various signal-to-noise ratios (SNR), as shown in Fig. \ref{fig:rob}, using a 1\% sampling ratio for all test cases. TiRE-GAN consistently outperforms the other methods. Moreover, its MSE growth is slow, particularly compared to advanced learning methods like RadioUNet and cGAN, which further demonstrates robustness of TiRE-GAN against noise.

 \vspace{-3mm}
\subsection{Outage Evaluation of Generated radiomap}
\begin{figure}[t]
    \centering
    \includegraphics[width=0.45\textwidth]{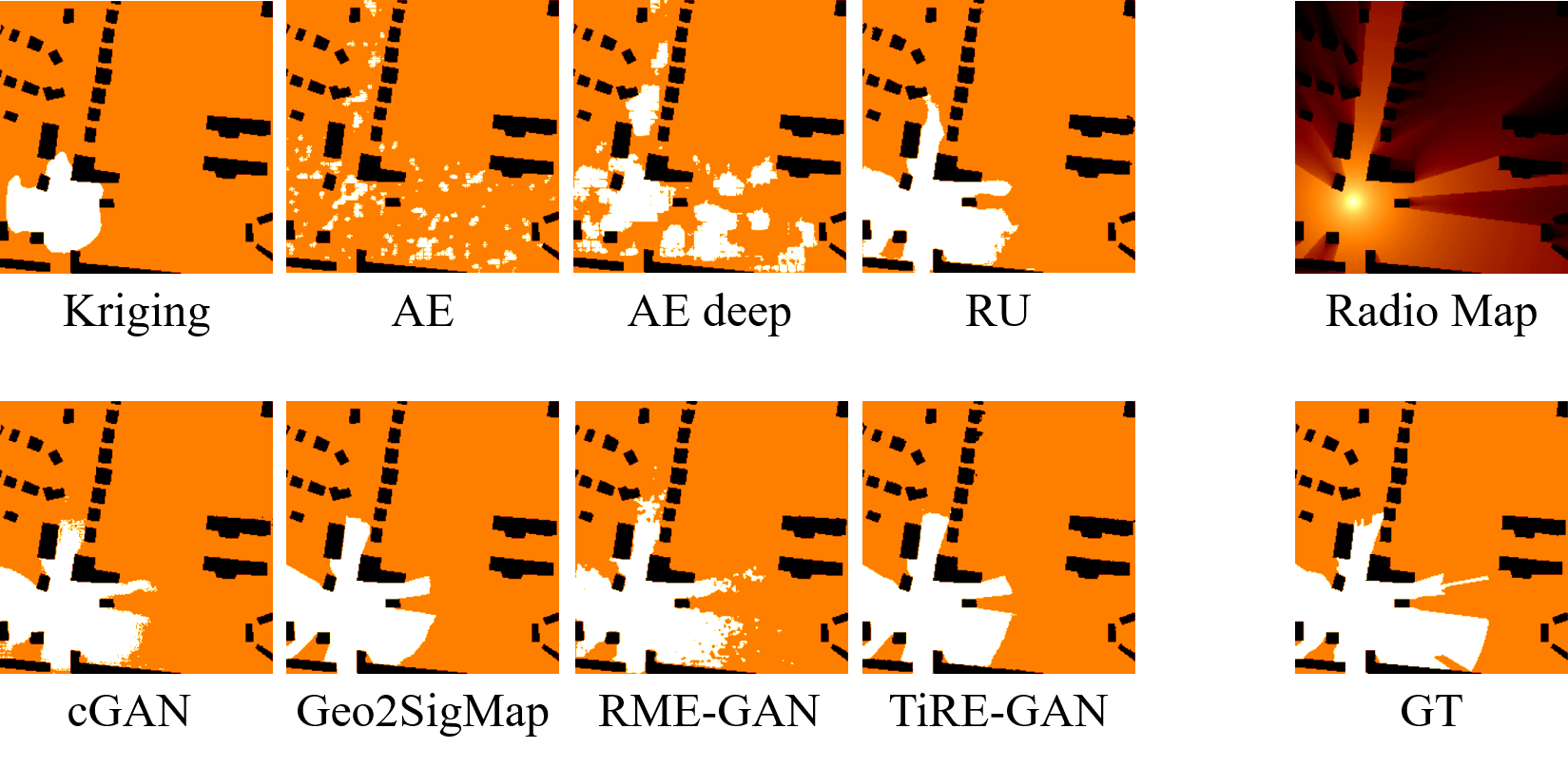}
    \vspace{-3mm}
    \caption{Illustration of Outage Map Prediction under sr=1\%. We compared the outage map predicted by the generated radiomap from different methods. The corresponding radiomap and the ground truth outage map(GT) are shown on the right for reference. }
    \label{fig:otg}
     \vspace{-2mm}
\end{figure}

To further evaluate the effect on downstream tasks, we compare the performance of RM-assisted outage detection. Here, we input the generated radiomaps from different methods into the pre-trained outage detection networks. The numerical results of the whole datasets with 1\% sampling and the exemplary visualization results are shown in Table \ref{tab:otg} and Fig. \ref{fig:otg}, respectively. {RME-GAN and TiRE-GAN achieve lower MSE than other methods, resulted from their utilization of model information. However,
in the visualization, TiRE-GAN displays smoother and more consistent patterns, which should benefit the spectrum management. These results demonstrate the effectiveness of TIN in radiomap compensation.}

\begin{table}[t]
\vspace*{-2mm}
    \centering
     \caption{Outage Map MSE comparison}
    \begin{tabular}{|c|c|c|c|c|}
    \hline
     Method  & Kriging &	AE  & AEdeep&RU  \\
     \hline
     MSE & 0.0781&0.0758	&0.0494 & 0.0284	 \\
     \hline
     Method  & cGAN& Geo2SigMap& RME-GAN&TiRE-GAN \\
     \hline
     MSE & 	0.0304	& 0.0561 & \textbf{0.0231}&\underline{0.0267} \\
     \hline
    \end{tabular}
    \label{tab:otg}
\vspace*{-3mm}
\end{table}

\vspace{-3mm}
\section{Conclusion}
TiRE-GAN is a task-incentivized generative learning model for radiomap estimation. To capture the overall radio propagation behavior, it uses a radio depth map as a physics-embedded feature to guide the data-driven generative model. Additionally, a task-incentivized network provides feedback to enhance detailed radiomap features for downstream tasks. Experimental results show that TiRE-GAN consistently outperforms other methods, especially with limited training samples, a faced by realistic applications. Future work will focus on optimizing the generator's reward mechanism and exploring multi-band radiomap estimation across different frequencies.



\end{document}